# Purely in-plane ferroelectricity in monolayer SnS at room temperature

Naoki Higashitarumizu[1], Hayami Kawamoto[1], Chien-Ju Lee[2], Bo-Han Lin[2], Fu-Hsien Chu[2], Itsuki Yonemori[3], Tomonori Nishimura[1], Katsunori Wakabayashi[3], Wen-Hao Chang[2,4], Kosuke Nagashio[1*]

[1]Department of Materials Engineering, The University of Tokyo, 7-3-1 Hongo, Bunkyo, Tokyo 113-8656, Japan

[2]Department of Electrophysics, National Chiao Tung University, Hsinchu 30010, Taiwan

[3]Department of Nanotechnology for Sustainable Energy, School of Science and Technology, Kwansei Gakuin University, Gakuen 2-1, Sanda, Hyogo 669-1337, Japan

[4]Center for Emergent Functional Matter Science (CEFMS), National Chiao Tung University, Hsinchu 30010, Taiwan

[*]nagashio@material.t.u-tokyo.ac.jp

**Abstract**

2D van der Waals ferroelectric semiconductors have emerged as an attractive building block with immense potential to provide multifunctionality in nanoelectronics. Although several accomplishments have been reported in ferroelectric resistive switching for out-of-plane 2D ferroelectrics down to the monolayer, a purely in-plane ferroelectric has not been experimentally validated at the monolayer thickness. Herein, a micrometer-size monolayer SnS is grown on mica by physical vapor deposition, and in-plane ferroelectric switching is demonstrated with a two-terminal device at room temperature (RT). SnS has been commonly regarded to exhibit the odd–even effect, where the centrosymmetry breaks only in the odd-number layers to exhibit ferroelectricity. Remarkably, however, a robust RT ferroelectricity exists in SnS below a critical thickness of 15 layers with both an odd and even number of layers. The lack of the odd–even effect probably originates from the interaction with the mica substrate, suggesting the possibility of controlling the stacking sequence of multilayer SnS, going beyond the limit of ferroelectricity in the monolayer. This work will pave the way for nanoscale ferroelectric applications based on SnS as a new platform for in-plane ferroelectrics.

Nanoscale ferroelectrics have been explored for decades in areas such as nonvolatile memories, sensors, and nonlinear optoelectronics. For 3D ferroelectrics, only a few successes have been reported in downscaling the film thickness: ~1 nm $BaTiO_3$[1,2] and 1.2 nm $PbTiO_3$[3]. Otherwise, ferroelectricity disappears in the nanoscale owing to the depolarization field or interfacial effects[4–6]. In contrast to the 3D materials, 2D layered materials have a dangling-bond-free surface with van der Waals (vdW) gap, and hence, they maintain the intrinsic properties even at an ultrathin thickness. Recent intensive works on 2D ferroelectrics[7–20] have experimentally demonstrated stable ferroelectricity down to the ultimate monolayer thickness for out-of-plane ferroelectrics ($MoTe_2$[7] and $WTe_2$[8]) and in-plane/out-of-plane intercorrelated ferroelectrics ($\alpha$-$In_2Se_3$[13,14] and SnTe[15]). Even in 2D ferroelectrics, however, the spontaneous polarization is also degraded with decreasing thickness when the 2D ferroelectric layer is vertically sandwiched with metals, owing to the depolarization field at metal/ferroelectric interfaces[21], as in 3D ferroelectrics. Therefore, in-plane ferroelectrics are superior to the out-of-plane and intercorrelated ferroelectrics in terms of preventing the depolarization field, given that the in-plane device structure enables a large gap between the electrodes.

The orthorhombic group-IV monochalcogenide (MX; M = Sn/Ge and X = S/Se), a purely in-plane 2D ferroelectric, has attracted considerable interest[22–27] because ferroelasticity and ferroelectricity have been predicted as multiferroicity with a larger spontaneous polarization ($P_s$ = 1.81 – 4.84 × $10^{-10}$ C/m)[22,23], compared with the above-mentioned 2D ferroelectrics. Moreover, the existence of spontaneous polarization guarantees piezoelectricity, and a remarkable piezoelectric coefficient of $d \sim 75–251$ pm/V has also been predicted[28], which is much larger than that of MoS$_2$ ($d \sim 4$ pm/V)[29] and comparable to that of Pb(Zr$_x$Ti$_{1-x}$)O$_3$ ($d \sim 300$ pm/V). These properties dominantly originate from a puckered structure along the armchair direction as a strong anisotropy analogue of black phosphorus. Given that MX is not a typical insulator for 3D ferroelectrics but a semiconductor, these properties will provide multifunctionalities in nanoscale devices. Among the MXs, SnS is the best because SnX is more chemically stable than GeX[30,31] and the Curie temperature of SnS is higher than that of SnSe[27]. Very recently, Bao *et al.* reported a ferroelectric device of bulk SnS (~15 nm), breaking the centrosymmetry by applying an external electric field[32]. As the few-to-monolayer SnS has been investigated only by piezoresponse force microscopy (PFM) owing to its small size of several tens of nanometers, the demonstration of a ferroelectric device for monolayer SnS has been challenging. Ferroelectricity in SnS has the odd–even effect owing to the stacking sequence of the centrosymmetric AB staking, as shown in **Fig. 1a**. The centrosymmetry exists in the even-number layers so that ferroelectricity is expected only in the odd-number layers and becomes prominent in few-to-monolayer SnS[22,26,32]. However, the synthesis of a high-quality monolayer SnS in the micrometer-size scale suitable for device fabrication has not been achieved[33–35], because the interlayer interaction is strong due to the lone pair electrons in the Sn atoms, which generate a large electron distribution and electronic coupling between adjacent layers[36,37].

Here, we report an in-plane ferroelectric device of a micrometer-size monolayer SnS grown by physical vapor deposition (PVD), where the growth conditions are precisely controlled to balance the adsorption/desorption of SnS. The Raman spectrum for monolayer SnS indicates high crystalline quality and strong anisotropy. Second harmonic generation (SHG) spectroscopy reveals that, unlike bulk SnS, monolayer SnS is non-centrosymmetric. Ferroelectric switching is successfully demonstrated for the monolayer device at room temperature (RT). Remarkably, for thin SnS below a critical thickness (~15 layers, L), the SHG signal and ferroelectric switching are also observed in the even-number SnS, thus overcoming the odd–even effect, which suggests that ultrathin SnS is grown in an unusual stacking sequence lacking centrosymmetry.

**Results**

**Growth of few-to-monolayer SnS.** SnS has the puckered structure along the armchair direction distorted from the NaCl structure (**Figs. 1a** and **b**), leaving the lone pair electrons in the Sn atoms. The lone pair electrons contributes to the strong interlayer force[36,37]. Therefore, the *in-situ* observation of SnS growth has confirmed a very high growth rate in the perpendicular direction[38]. Thus, monolayer SnS has been realized by molecular beam epitaxy growth, although only with a limited crystal size of several tens of nanometers[32]. Otherwise, the minimum thickness was 5.5 nm *via* PVD growth[34]. To suppress the perpendicular growth rate, SnS desorption during the PVD growth was precisely controlled with growth pressure and temperature (Supplementary Fig. 1). Consequently, a thickness controllable PVD process was realized from bulk to monolayer thickness (**Fig. 1c** and Supplementary Fig. 2). **Fig. 1d** and Supplementary Fig. 3 show atomic force microscopy (AFM) topographic images of SnS grown on mica substrates. For bulk SnS thicker than ~16 nm, the SnS crystal has a sharply defined diamond shape that reflects the thermodynamically stable crystal facets (**Fig. 1b**). With decreasing temperatures of the SnS source powder ($T_{source}$) and substrate ($T_{sub}$), the SnS thickness decreased and the corner became rounded. The typical temperatures ($T_{source}$, $T_{sub}$) for the SnS crystals with thicknesses of ~36 nm and ~3-4 nm were (590°C, 530°C) and (530°C, 410°C), respectively. Finally, at $T_{source}$ = 470°C and $T_{sub}$ =

410°C, monolayer-thick SnS was realized with micrometer size of up to ~5 µm, which is a reasonable size for the device fabrication. Note that monolayer-thick SnS has an atomically flat surface without wedding cake morphology owing to the Stranski–Krastanov growth mode[32] or spiral growth assisted by a screw dislocation[38]. The rounded shape could be caused by the SnS desorption during the growth and an insufficient growth time to reach the thermodynamic equilibrium state for thin SnS[35,38]. The small "holes" found in the AFM image of monolayer-thick SnS are probably etch pits created during the growth. When a post-growth annealing was performed for the SnS crystals, aligned square-shaped etch pits were created, indicating a single crystalline nature (Supplementary Fig. 4). Moreover, a lattice matching of SnS with the mica substrate was detected by the in-plane X-ray diffraction (XRD) measurement, suggesting a strong SnS/mica interaction (Supplementary Fig. 5)[39].

**Characterization of few-to-monolayer SnS. Fig. 1e** shows a cross-sectional bright-field scanning transmission electron microscopy (STEM) image of trilayer SnS. For trilayer SnS, the composition ratio Sn : S was 1 : 0.8 from energy-dispersive X-ray spectroscopy (EDS) (Supplementary Fig. 6), indicating the absence of other S rich phases (e.g., $SnS_2$ and $Sn_2S_3$). Interestingly, from the TEM image of trilayer SnS, the monolayer thickness $d_{1L}$ was determined to be 5.8 Å, larger than that of the mechanically exfoliated bulk sample ($d_{1L}$ ~5.4 Å[37]). This expansion suggests the possibility of an unusual stacking sequence rather than the AB stacking. From an ab initio simulation, the AA stacked bilayer SnS (**Fig. 1a**) indicates $d_{1L}$ = 6.34 Å, larger than $d_{1L}$ = 5.85 Å for the AB stacking. For the TEM observation of monolayer SnS, the crystal orientation was determined beforehand by polarized Raman spectroscopy, as discussed later, because the TEM image of monolayer SnS is much more indistinct than that of the bulk crystals owing to the degradation during the sample preparation by the focused ion beam (FIB) process and the TEM observation itself. By adjusting the zone axis, it can be observed that two sub-layers have a monolayer structure, which matches well with the configuration along the armchair direction (**Fig. 1f**).

**Fig. 1g** shows Raman spectra for PVD grown SnS with different thicknesses from bulk to monolayer, measured at 3K. For a bulk SnS (~37 L), specific peaks were observed at 153.7, 188.3, 227.6, and 291.8 cm$^{-1}$ in addition to the peaks from mica substrate at ~131, 198.9, 276.0, 312.8, 322.9, and 372.5 cm$^{-1}$. These peak positions are well consistent with those of bulk SnS *via* mechanical exfoliation or PVD[33–35,37]. With a decreased thickness from trilayer to monolayer, the Raman peak positions at approximately 230 and 293 cm$^{-1}$ almost coincided with each other, while those between 140 and 190 cm$^{-1}$ changed significantly. For monolayer SnS, a Raman peak was observed at ~145 cm$^{-1}$, and also at 232.4 and 295.0 cm$^{-1}$. A similar trend was also obtained at RT (Supplementary Fig. 7). Those values will be discussed later with the results of polarized Raman spectroscopy and the phonon mode calculation. To determine the non-centrosymmetry, which is required for ferroelectricity, μ-SHG spectroscopy[40,41] was carried out for different thicknesses, from bulk to monolayer (**Fig. 1h**). An 850-nm laser was used as the excitation source. The bulk SnS, thicker than ~21L, showed no SHG signal, while SnS under the critical thickness of ~15L showed SHG signal at $\lambda = 425$ nm. Although the odd–even effect was expected for the AB stacked SnS, it was found that all the ultrathin SnS flakes, including the even-number layers, showed SHG signals below the critical thickness (Supplementary Fig. 8). For confirmation, mica substrate was measured under the same conditions. Although it is known that a weak SHG signal can be generated even for the centrosymmetric material due to the surface SHG effect[40], no SHG signal was detected from the mica surface (**Fig. 1h**). These results suggest an unusual stacking sequence of the PVD grown SnS, along with the results of interlayer distance from the TEM image.

**Anisotropic characteristics for monolayer SnS.** In monolayer SnS, the dipole moment along the armchair direction leads to spontaneous polarization[22,23]. Different from the thick SnS, it is difficult to identify the orientation of the present few-to-monolayer crystals from their shape, as mentioned above. To characterize the in-plane anisotropy, the angular dependences of both Raman and SHG have been investigated. **Fig. 2a** shows the polarization dependence of the Raman spectrum for monolayer SnS under the parallel polarization configuration at 3 K (Supplementary Fig. 9). Specific peaks were observed at approximately 234 and 294 $cm^{-1}$, which are consistent with the results of the unpolarized Raman measurement, although it was difficult to determine the precise peak position between 100 and 200 $cm^{-1}$ because of the overlaps with peaks from the substrate. To investigate the relationship between the Raman active modes and stacking sequences for SnS, an ab initio calculation was carried out using the Vienna ab initio simulation package (VASP)[42]. **Fig. 3a** shows a typical example of phonon dispersion along the path passing through the main high-symmetry $k$-points in the irreducible Brillouin zone of monolayer SnS. **Fig. 3b** and Supplementary Fig. 10 summarize the Raman active phonon modes for the AB and AA stacked SnS with different number of layers. For bulk SnS, the experimental results almost agree with the calculated results, whereas there are large differences between experiments and calculation for the few-to-monolayer SnS. The origin of this difference is probably the strain incorporated through the interaction with the mica substrate, as discussed above. By comparing the calculated Raman active modes and experimental results, the Raman signals of the monolayer at 234 $cm^{-1}$ is attributed to the $A_1$ mode of the $C_{2v}$ point group (Supplementary Fig. 10). As expected, the Raman peak intensity at 234 $cm^{-}$ [1] shows a significant change as a function of the rotation angle (**Fig. 2a**). The Raman tensor $R$ for the $A_1$ mode of $C_{2v}$ point group can be written as[43]

$$R(A_1) = \begin{pmatrix} |A|e^{i\varphi_A} & 0 & 0 \\ 0 & |B|e^{i\varphi_B} & 0 \\ 0 & 0 & |C|e^{i\varphi_C} \end{pmatrix} \tag{1}$$

which is the same as the $A_g$ mode of bulk SnS ($D_{2h}$ point group) that shows a strong anisotropy[33–35]. The unitary vector of incident light is $e_i = (\cos\theta, \sin\theta, 0)$, where $\theta$ is the polarization angle defined as the angle between the incident light and the zigzag direction of SnS crystal. The unitary vector of scattered light is $e_s = (\cos\theta, \sin\theta, 0)$ and $(-\sin\theta, \cos\theta, 0)$ for the parallel (∥) and perpendicular (⊥) polarization, respectively. For the polarized Raman intensity of the $A_1$ peak, the angular dependences can be calculated using the following equations[35]:

$$I_\parallel \propto |A|^2 \cos^4\theta + |B|^2 \sin^4\theta + 2|A||B|\cos^2\theta \sin^2\theta \cos\varphi_{BA} \tag{2}$$

$$I_\perp \propto \frac{|A|^2 + |B|^2 - 2|A||B|\cos\varphi_{BA}}{4} \sin^2 2\theta \tag{3}$$

where $\varphi_{BA} = |\varphi_B - \varphi_A|$ is the phase difference between the Raman tensor elements $|A|e^{i\varphi_A}$ and $|B|e^{i\varphi_B}$. By fitting the experimental data with **Eq. (2)**, the crystal orientation was revealed (**Fig. 2c**). As in the polarized Raman spectroscopy, **Fig. 2b** shows a strong angular dependence of SHG for monolayer SnS with perpendicular polarization configuration. For the polarized SHG intensity under parallel and perpendicular polarization, the angular dependence in the $C_{2v}$ point group is written as[44]

$$\chi_\parallel^{(2)} = \left(\chi_{xyx}^{(2)} + \chi_{yxx}^{(2)}\right)\sin\theta\cos^2\theta + \chi_{yyy}^{(2)}\sin^3\theta \tag{4}$$

$$\chi_\perp^{(2)} = \chi_{yxx}^{(2)}\cos^3\theta + \left(\chi_{yyy}^{(2)} - \chi_{xyx}^{(2)}\right)\cos\theta\sin^2\theta \tag{5}$$

where $\chi_{ijk}^{(2)}$ is the SHG susceptibility tensor element along the different directions. We fitted the experimental data based on **Eq. (5)** to determine the zigzag/armchair orientation, as shown in **Fig.**

**2d**. The measured patterns agree well with the theoretical model. The anisotropy revealed from the polarized Raman and SHG spectra indicates again the high crystallinity of monolayer SnS.

**Modulation of Schottky barrier height at metal/SnS interface.** Switching of the spontaneous polarization and SHG is required to prove the ferroelectricity in SnS. A tip poling experiment by using scanning probe microscopy is an effective way to observe the polarization switching for the local area. However, it is more difficult to detect the trace of spontaneous polarization in SnS than that in out-of-plane ferroelectrics because in-plane ferroelectricity does not respond to the out-of-plane electric field applied by the probe tip. To demonstrate the in-plane polarization switching, in-plane two-terminal devices with source/drain electrodes on SnS crystals were fabricated (**Fig. 4a** and Supplementary Fig. 12). During the fabrication process, the exact locations of SnS crystals were captured by optical images so that only one crystal were contacted by two adjacent electrodes. Note again that SnS is a semiconductor with energy gaps of 1.5 eV for monolayer and 1.1 eV for bulk[45], where channel conductance makes it difficult to identify the very small displacement current[46]. To prevent the channel conductance by forming an insulator-like interface, the Schottky barrier height (SBH) has been investigated by changing the metal work function ($\Phi_\mathrm{m}$). After bulk SnS (~20 nm) was grown on the mica substrate, a standard electron beam lithography was performed followed by multiple metal depositions with a series of metals (In, Al, Ag, Cu, Ni, Pd, and Au), as shown in **Fig. 4a**. Assuming that the metal/SnS interface is ideal and free from Fermi level pinning[47], the SBH will strongly depend on $\Phi_\mathrm{m}$. In such a case, the metal with smaller $\Phi_\mathrm{m}$ is preferable to increase the SBH for the *p*-type semiconductor SnS[33,37,48,49]. At RT, ohmic $I_\mathrm{D}$–$V_\mathrm{D}$ curves were obtained for bulk SnS with metals of In, Cu, Ni, Pd, and Au, while Schottky $I_\mathrm{D}$–$V_\mathrm{D}$ with Al and Ag (**Fig. 4b**, and Supplementary Fig. 13).

**Ferroelectric switching device of few-to-monolayer SnS. Fig. 4c** shows $I_D$–$V_D$ curves for 9L SnS with the Ag contact measured by increasing the $V_D$ sweep range at RT. Unlike the $I_D$–$V_D$ curve without hysteresis for the ohmic Ni metal contact (Supplementary Fig. 14), the Schottky Ag contact exhibited a well-reproducible hysteresis (Supplementary Fig. 15). Although it is known that Ag could give memristor behavior owing to its high diffusivity[50,51], no significant no Ag diffusion was confirmed from the AFM images at the low resistive state (Supplementary Fig. 16). Moreover, the larger drain bias led to a larger window of hysteresis loop, distinguishing a maximum conductivity at approximately $V_D = \pm 1$ V. To confirm that this hysteresis originates from the ferroelectric switching, a double-wave measurement was performed[52]. When $V_D$ was applied from 0 to 2 V for two times, a current peak was observed only in the first sweep and it turned to be a highly resistive state in the second sweep (**Fig. 4d**). Similarly, a negative $V_D$ sweep (0 to −2 V) showed a current peak only once in the first sweep. This is because polar switching results in the current peak in the first sweep, but never in the second sweep because the direction of polarization is steady. Therefore, this result of double-wave measurement is strong evidence for the switching of spontaneous polarization, that is, ferroelectricity. It should be emphasized that that the crystal orientation of SnS was not determined before the electrode fabrication. Despite the fact that spontaneous polarization exists along the armchair direction, the $I_D$–$V_D$ hysteresis loop was observed regardless of the initial crystal orientation that the external electric field is applied. This result suggests that the more flexible polarization switching due to the possible existence of the domain structure. Furthermore, even an external electric field along the non-ferroelectric zigzag direction could induce ferroelectricity with structural rearrangement into the armchair structure[53] in a similar way as ferroelasticity[22,23].

To further analyze the switching behavior quantitatively, $Q$–$E$ ($Q$ is charge, and $E$ is electric field) curve was measured using a ferroelectric evaluation system. When the AC bias at 1 Hz was applied (0 → +2 → −2 → 0 V), an $I$–$E$ curve drew a hysteresis loop very similar to the $I_D$–$V_D$ curve under DC bias (**Fig. 4e**). Here, $E$ is assumed to be $V_D/l_{ch}$, where $l_{ch}$ is the channel length, which should overestimate $E$ because there is also a voltage drop at the Schottky contact. The $Q$–$E$ loop corresponding to the $I$–$E$ loop is shown at the bottom of **Fig. 4e**. A distinct hysteresis loop was obtained with the charge of ~9 pC, which is a characteristic of ferroelectrics in general when the external electric field switches the polarization to generate the displacement current. However, the remnant polarization $P_r = Q/w_{ch}$, where $w_{ch}$ is channel width, was determined to be $P_r \sim 3$ μC/m, which is almost four orders of magnitude larger than the theoretical value of $P_r = 260$ pC/m[23]. This discrepancy suggests that the hysteresis loop is not mainly due to the displacement current. The hysteresis loop can be dominantly caused by the contribution of other factors in addition to the displacement current, as discussed below.

Considering it has been suggestively pointed out that the closed $Q$–$E$ loop can be artificially obtained even for non-ferroelectric materials when they are lossy or leaky[54,55], the large discrepancy on the remnant polarization for SnS is carefully discussed as follows. One possible case is for films with a large concentration of traps near the metal/film interface[55]. The number of traps required to reproduce the $Q$–$E$ loop in **Fig. 4e** was estimated. As it corresponds to one in three of the total number of atoms in the SnS channel material, the traps are not the origin for the large $P$ (Supplementary Fig. 17). The other case is the resistive switching[56], where a hysteresis loop similar to that in **Fig. 4e** has been discussed as an effect of SBH modulation at the metal/ferroelectric interface owing to the reversed polarization for the previous works on ferroelectric 2D materials (MoTe$_2$[7], WTe$_2$[8], α-In$_2$Se$_3$[13,14,18,19], SnTe[15], and CuInP$_2$S$_6$[16,17]). In the present study, the current flowing through the SnS channel was drastically suppressed by selecting the strong Schottky contact metal. Nevertheless, the present ferroelectric $Q$–$E$ hysteresis loop is considered to be owing to the resistive change at the metal/SnS Schottky interface accompanied with polar switching because the displacement current level for in-plane 2D devices is negligibly small. Moreover, the coercive electric field $E_c$, which originates this switching, was found to be ~25 kV/cm for 9L SnS. This value is comparable to the experimental value for bulk SnS with gate-induced non-centrosymmetry (~10.7 kV/cm)[32], whereas it is much smaller than the theoretical value for monolayer ($1.8 \times 10^3$ kV/cm)[24]. This discrepancy between the experiment and calculation is probably related to the existence of a mobile domain wall or lattice strain in SnS caused by the difference in thermal expansion coefficients between SnS and the mica substrate in the real system[23].

## Discussion

For further understanding the dependences of SHG and ferroelectric switching on the number of layers, the SnS thickness was systematically changed. **Fig. 5a** shows $I_D$–$V_D$ curves measured in the same way as that in **Fig. 4c**. For the monolayer, bilayer, and trilayer, the ferroelectric switching was realized as in 9L SnS, as discussed above. For SnS thicker than 15L, the current leakage through the SnS channel dramatically increased and it was difficult to observe the ferroelectric hysteresis loop. To quantify the effect of polar switching on the resistivity, the conductivity ratio for the low resistive state (ON) and high resistive state (OFF) at $E_c$ was calculated. **Fig. 5b** shows the dependence of the ON/OFF ratio on the number of layers together with the SHG intensity. A large distribution was found in the SHG intensity for each number of layers, which is probably caused by the variation of crystalline quality due to the desorption-controlled PVD growth. This heterogeneity was also found when a different batch of SnS powder source was used. That is, the maximum SHG intensity for SnS grown via one batch exceeded that grown via the other batch. This result suggests the possibility of further increase in the SnS crystalline quality with the improvement of the SnS powder source. Despite of these dispersions, the SHG intensity tends to increase with a number of layers up to ~10L, then it decreases, and finally quenches above 21L. The ferroelectric switching only occurs below this critical thickness. These results indicate that there is a change in the stacking sequence from the AA to AB stacking, which determines the existence of ferroelectricity (Supplementary Fig. 19). In order to prove the stacking transition, cross-sectional structure was further investigated for 16L SnS by TEM observation. For this sample, the crystal orientation was determined from the diamond-shaped-like crystal structure so that we can observe the atomic configuration along the armchair direction, where AA and AB stackings can be identified (**Fig. 5c** and Supplementary Fig. 20). **Fig. 5c** shows high-angle annular dark-field (HAADF) STEM image of 16L SnS along the armchair direction. The stacking transition from AA to AB is clearly observed at the thickness of 6L. Even though this transition plane was steeply continuous through a selected region of several tens of nanometers, it can be possible that the transitions occur at different thicknesses in a micrometer scale and from sample to sample, resulting in the discrepancy between critical thicknesses determined from TEM and SHG measurements. The unusual growth mode is probably due to the substrate effect, such as lattice strain and electrostatic surface charges. In the previous work on bulk SnS (~15 nm)[32], an $I_D$–$V_D$ hysteresis loop similar to that in the present work was observed for in-plane two-terminal Au contact devices, where the ferroelectricity in bulk SnS was achieved by extrinsically breaking the inversion symmetry through the perpendicular electric field from the back gate. It should be emphasized that the intrinsic ferroelectricity is observed for monolayer SnS in this study.

In conclusion, monolayer SnS with micrometer size is grown by precisely controlling the growth pressure and temperature in PVD. The lack of the centrosymmetry and a strong anisotropy rooted from the puckered structure is confirmed based on polarized Raman and SHG spectroscopies. After the current flowing through the SnS channel in two-terminal devices was suppressed by selecting the strong Schottky Ag contact, RT in-plane ferroelectric switching was realized by the double-wave method. Remarkably, the robust RT ferroelectricity was identified in SnS below the critical thickness of ~15L, probably due to the interaction with the substrate. This result suggests a possibility of controlling the stacking sequence of multilayer SnS, going beyond the limit of ferroelectricity in monolayer SnS. Given that SnS is the semiconductor with multiferroicity[22,23], innately exhibiting pyroelectricity and piezoelectricity, this work will open up possibilities of providing novel multifunctionalities in vdW heterostructure devices.

**Methods**

**PVD growth.** SnS crystals were grown by a home-built PVD growth furnace with three heating zones (Supplementary Fig. 1). A commercially available SnS powder was used as a source. To promote lateral growth, we used a freshly cleaved mica substrate sized 1 cm × 1 cm × 0.5 mm, whose surface is atomically flat. $N_2$ carrier gas was introduced into the furnace through the mass flow controller and the growth pressure was reduced to 10 Pa by a vacuum system to enhance the SnS desorption during the growth.

**Optical characterizations.** μ-Raman spectra were measured using a 488 nm excitation laser, whose penetration depth is ~20 nm in SnS. The nominal $1/e^2$ spot diameter and laser power on the sample surface were 2.5 μm and 0.5 mW, respectively. To avoid degradation of SnS during the measurement, the samples were measured in the vacuum. The SHG measurements were conducted using a mode-locking Ti:sapphire laser (wavelength: 850 nm, pulse width: ~150 fs and repetition rate: 80 MHz) in a home-built optical microscope under the backscattering configuration. The laser pulse was focused to a spot size ~1.1 μm on the sample by a 100× objective lens. The backscattered SHG signals were sent into a 0.75-m monochromator and detected by a nitrogen-cooled CCD camera. For polarization-resolved SHG, the sample was mounted on a motorized rotational stage. The linear polarization of the excitation laser and SHG signals was selected and analyzed separately by polarizers and half-wave plates.

**Ab initio Simulation.** We have used the VASP to perform first-principles calculations based on density functional theory to study geometric and electric properties[42]. The exchange and correlation potentials are the Perdew-Burke-Ernzerhof (PBE) functional and is treated using the generalized gradient approximation (GGA)[57]. We employed the Monkhorst-Pack scheme to sample reciprocal space with Γ-centered $16 \times 16 \times 1$ grid for geometry relaxations of 2D systems and $16 \times 16 \times 4$ grids for that of bulk system. The plane-wave basis cutoff energy is set to be 500 eV. The convergence criterion is set to be $10^{-5}$ eV for energy in SCF cycles. And the full relaxation is continued until the residual force is less than 0.01 eVÅ$^{-1}$. We set 25 Å vacuum perpendicular to the 2D plane is used to avoid the interaction between replaced atoms. In addition, Grimme's DFT-D2[58] method implemented in VASP is invoked to correct the vdW-like interaction existing in these systems. Moreover, we calculated phonon dispersion at Γ-point based on density functional perturbation theory. In addition, the calculation of phonon dispersion and irreducible representation has been implemented using phonopy[59].

**Device fabrication and transport characterization.** Two-terminal devices were fabricated with electrode pattering using standard electrode beam lithography. Electrode metals (In, Al, Ag, Cu,

Ni, Pd, and Au) were deposited via thermal evaporation, followed by additional Au deposition as passivation for the source/drain metals (Supplementary Fig. 12). The electrical transport characteristics were measured in vacuum to avoid sample degradation. The $Q$–$E$ curves were measured by a ferroelectric evaluation system (FCE-1A, Toyo corporation) with samples set in vacuum.

Acknowledgements

K. N. acknowledges the funding supports by The Murata Science Foundation, Maekawa Houonkai foundation, Yazaki memorial foundation for science and technology, Precision measurement technique foundation, The Canon Foundation, the JSPS Core-to-Core Program, A. Advanced Research Networks, the JSPS A3 Foresight Program, and JSPS KAKENHI Grant Numbers JP19H00755, and 19K21956, Japan. N. H. was supported by JSPS KAKENHI Grant-in-Aid for JSPS Fellows Number JP19J13579, Japan. W.H.C acknowledges the supports from the Ministry of Science and Technology of Taiwan (MOST-108-2119-M-009-011-MY3, MOST-107-2112-M-009-024-MY3) and from the CEFMS of NCTU supported by the Ministry of Education of Taiwan. K. W. acknowledges JSPS KAKENHI Grant Number JP 18H01154 and financial support from Masuya Memorial Research Foundation of Fundamental Research.

Author contributions

K.N. and N.H. conceived and designed the project. N.H. and H.K. performed PVD growth of SnS and Raman spectroscopy. W.H.C., C.J.L., B.H.L., and F.H.C. performed SHG measurements. K.W. and I.Y. performed ab initio Simulation. N.H. fabricated the devices and performed the electrical characterization. N.H. and T.N. performed ferroelectric measurements and XRD characterization. K.N. and N.H. analyzed the data and wrote the manuscript with input from all the authors.

Additional information

Competing interests: The authors declare no competing interests.

**Figures**

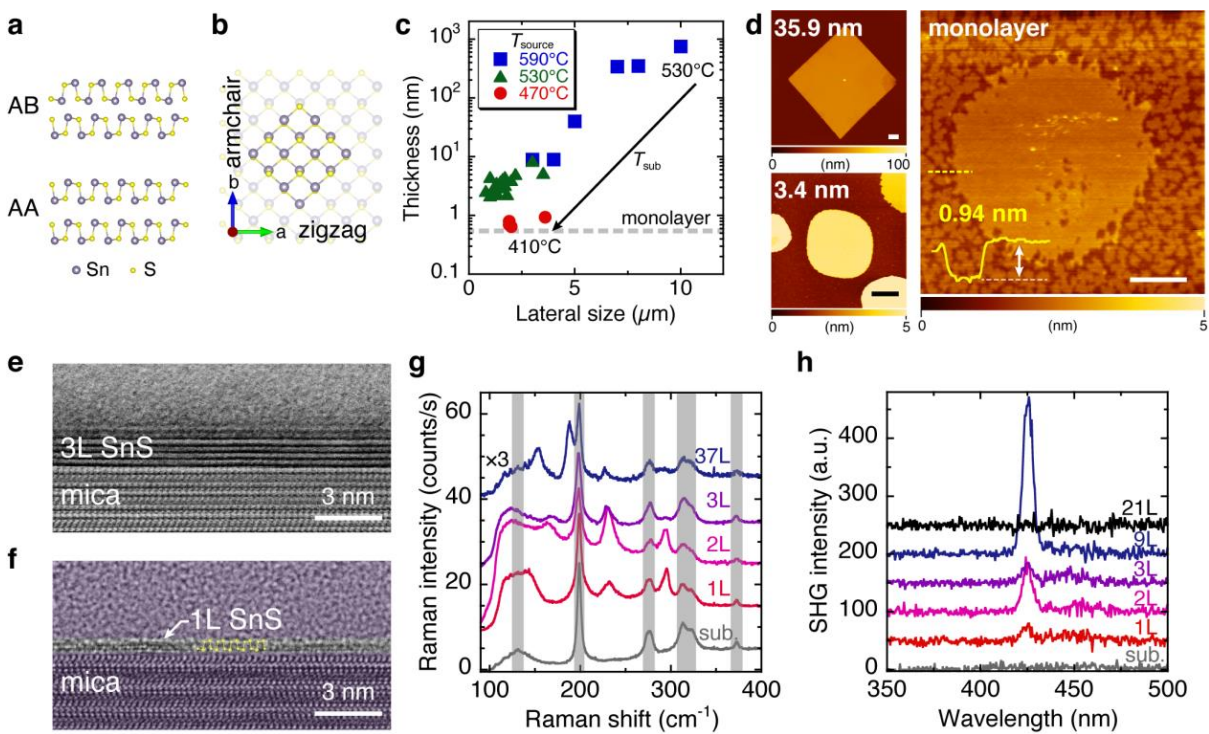

**Fig. 1** Characterization of few-to-monolayer SnS. **a** Cross-sectional crystal structures of SnS along the armchair direction with different stacking sequences: non-centrosymmetric AA and centrosymmetric AB staking. **b** Top view of crystal structure of monolayer SnS, whose two-fold axis is along the armchair direction. Highlighted area shows thermodynamically stable facets. **c** The minimum thickness versus lateral size of PVD grown SnS with changing $T_{source}$ and $T_{sub}$. **d** AFM topographic images of SnS crystals with different thicknesses from bulk to monolayer. The scale bars represent 1 µm. **e** Cross-sectional bright-field STEM image of trilayer SnS. **f** Cross-sectional TEM image of monolayer SnS along the armchair direction. As guide to the eye, all of the region except the SnS crystal is shaded, and the atomic model is overlaid on the TEM image. **g** Thickness dependence of Raman spectrum for SnS at 3 K. The peaks in the hatch come from the mica substrate. **h** SHG spectra for SnS with different thicknesses from bulk to monolayer at RT.

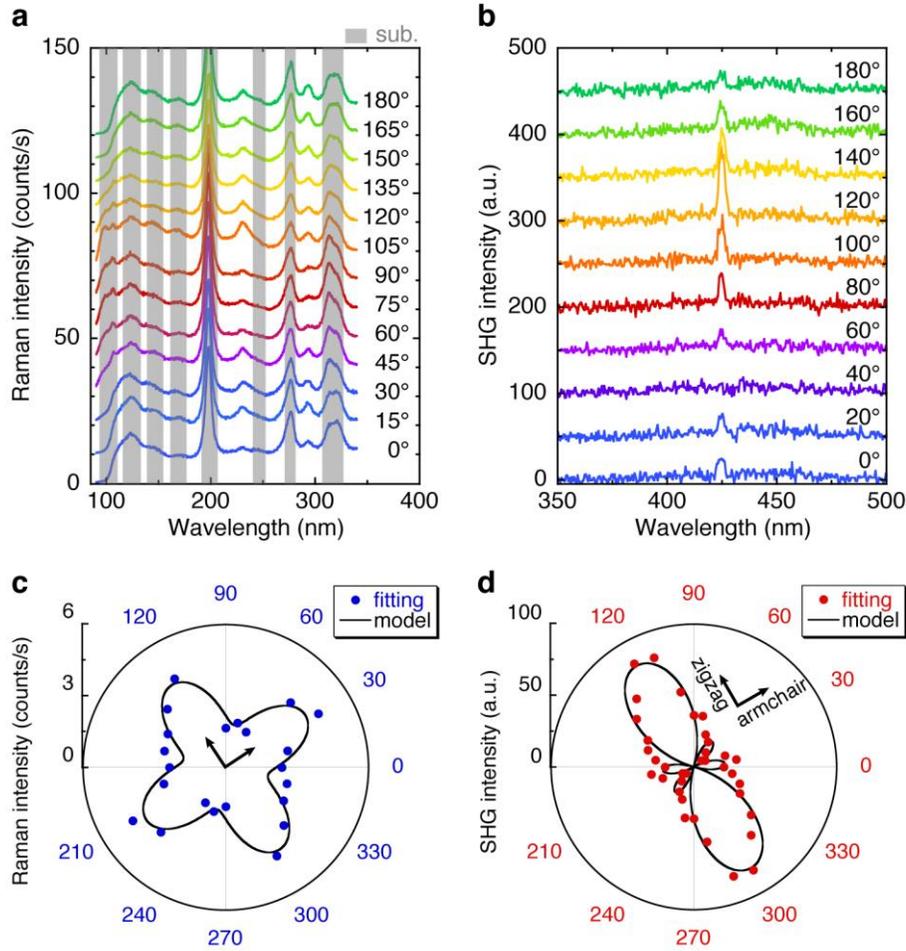

**Fig. 2** Optical anisotropies of monolayer SnS. **a,b** Polarization dependences of Raman (3 K) and SHG (RT) spectrum of monolayer SnS, with parallel and perpendicular polarization, respectively. The gray shaded region of Raman spectra represents Raman peak from mica substrate. **c,d** Polar plots of Raman intensity at ~234.0 cm$^{-1}$ and SHG intensity at 425 nm, respectively. The inset axes show the armchair and zigzag directions.

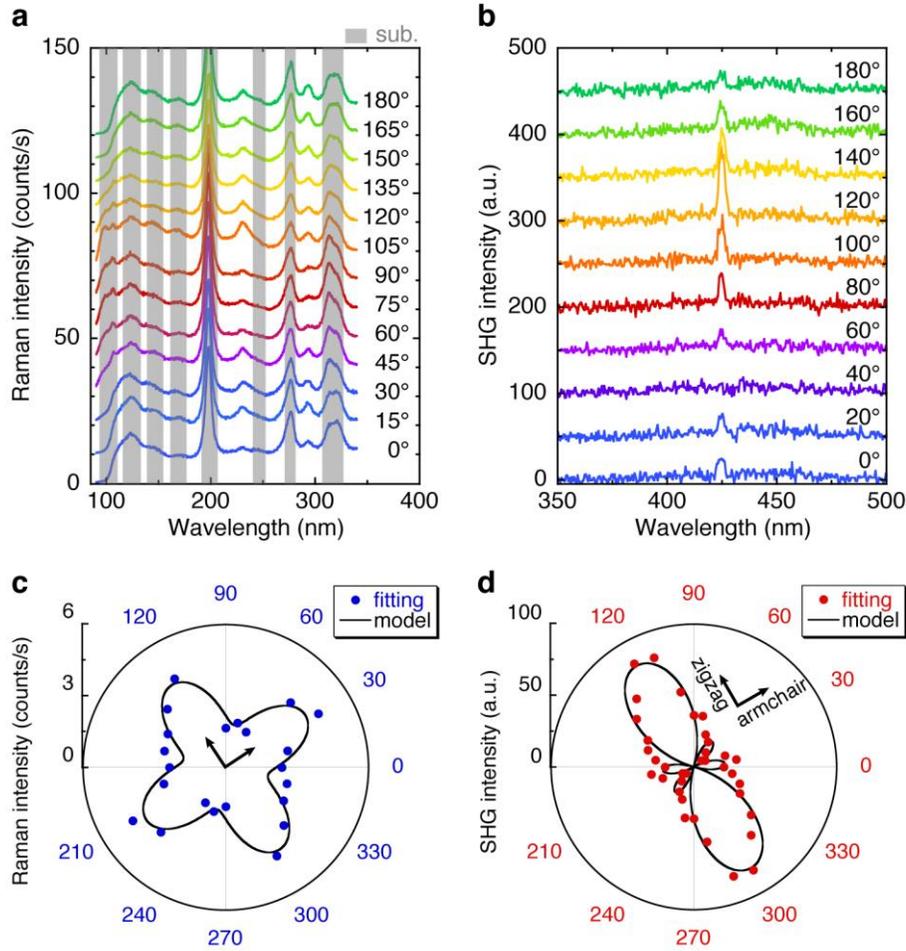

**Fig. 2** Optical anisotropies of monolayer SnS. **a,b** Polarization dependences of Raman (3 K) and SHG (RT) spectrum of monolayer SnS, with parallel and perpendicular polarization, respectively. The gray shaded region of Raman spectra represents Raman peak from mica substrate. **c,d** Polar plots of Raman intensity at ~234.0 cm$^{-1}$ and SHG intensity at 425 nm, respectively. The inset axes show the armchair and zigzag directions.

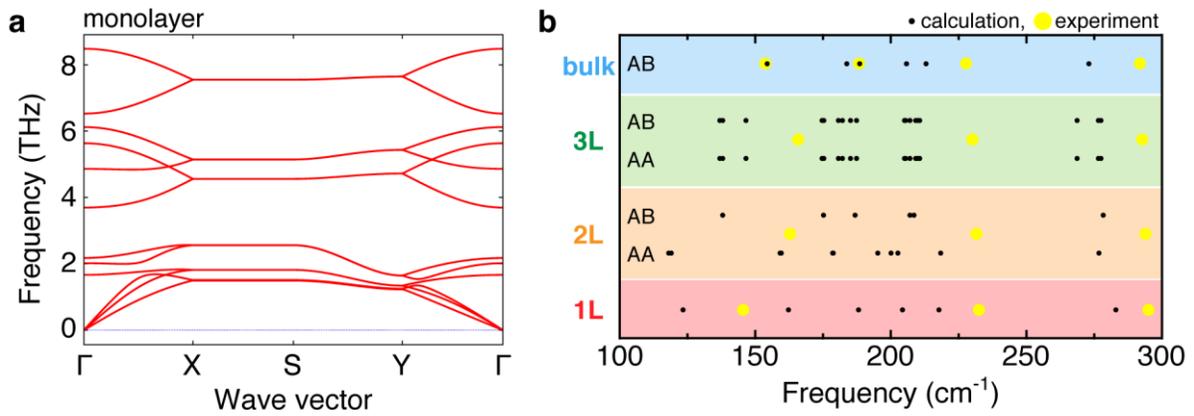

**Fig. 3** Theoretical calculation of phonon dispersion and Raman active modes. **a** Phonon dispersion of the monolayer SnS. **b** Comparison of calculated Raman active modes and experimental Raman peak positions for SnS with different thicknesses (monolayer, bilayer, trilayer, and bulk) and stacking sequences (AA and AB).

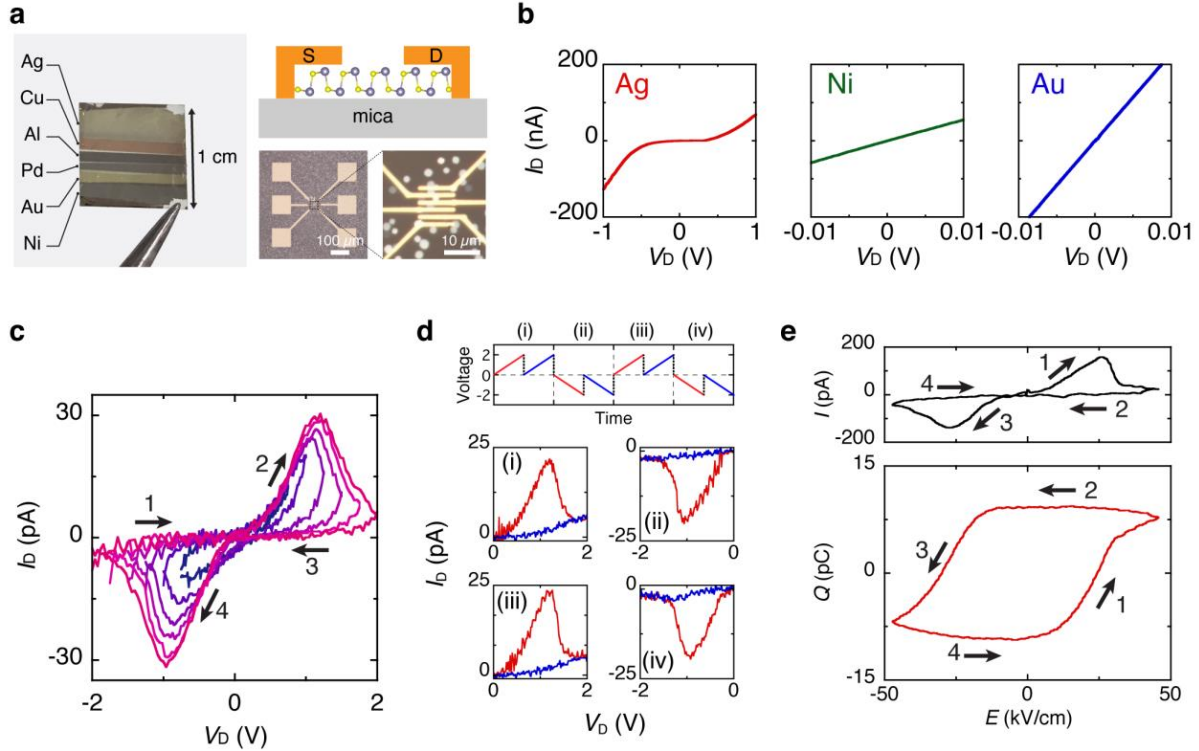

**Fig. 4** Ferroelectric switching behaviors of few-layer SnS. **a** Left: Photograph of mica substrate after the deposition of various metals with different work functions. Right: Cross-sectional schematic and optical images of two-terminal SnS devices. Typical channel length $l_{ch}$ and width $w_{ch}$ were $l_{ch}$ = 0.4–0.8 µm and $w_{ch}$ = 3–5 µm, respectively. **b** $I_D$–$V_D$ curves for bulk SnS with different metal contacts: Ag, Ni, and Au. **c** RT $I_D$–$V_D$ for Ag/9L-SnS device with different $V_D$ sweep ranges. $V_D$ was swept from minus to plus to minus (e.g., −1 → +1 → −1 V). **d** Double-wave measurement from 0 to 2 and 0 to −2 V. Top: applied voltage along time. The voltage was applied two times at the positive and negative bias repeatedly. Bottom: $I_D$–$V_D$ curves for different sweeps (i)–(iv). The red and blue lines represent the first and second sweep, respectively. **e** Ferroelectric resistive switching for Ag/SnS: current and charge versus nominal electric field measured by ferroelectric measurement system at 1 Hz and RT.

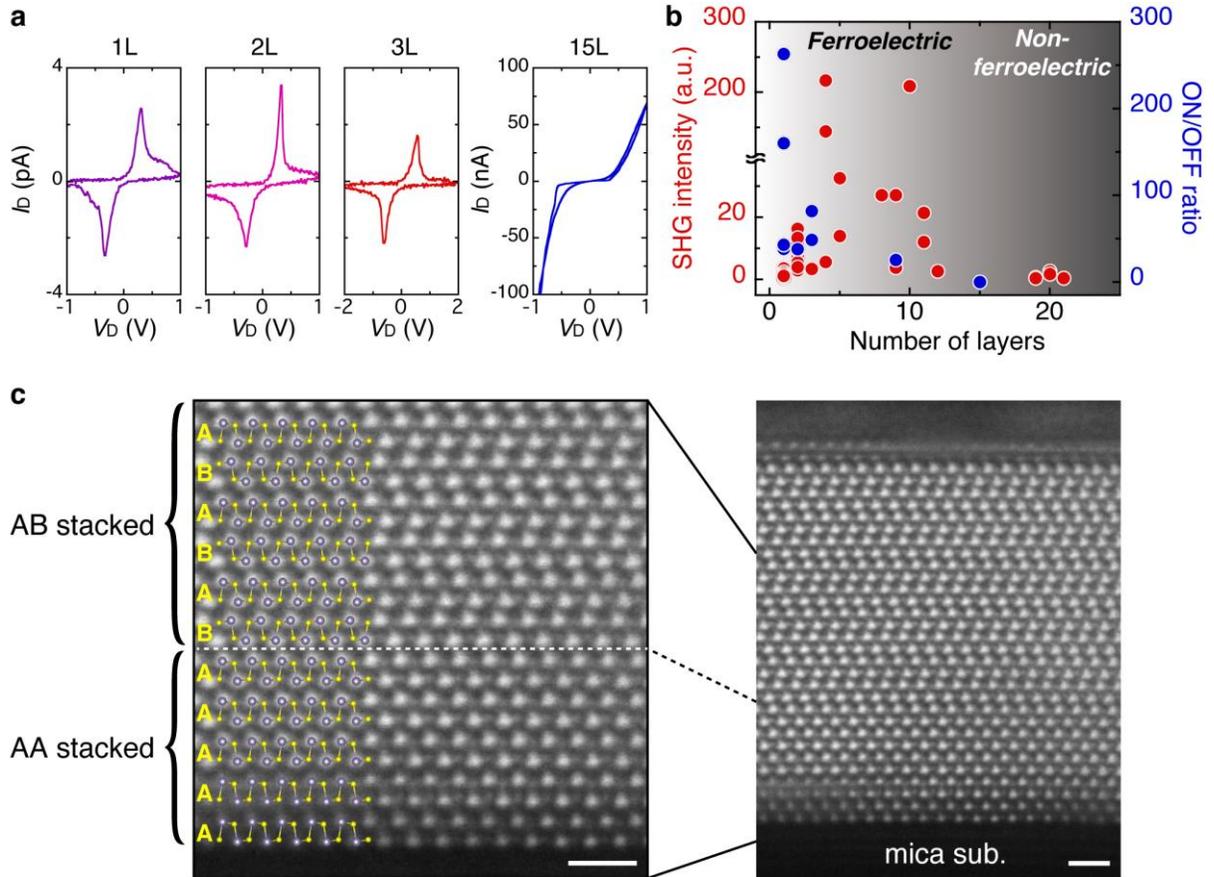

**Fig. 5** Transition of stacking sequence from AA to AB staking. **a** Ferroelectric resistive switching for SnS with different thicknesses: monolayer, bilayer, trilayer, and 15L. **b** Thickness dependences of the SHG intensity and ON/OFF ratio for different thicknesses. The ON/OFF ratio was determined at the coercive electric field of $I_D$–$V_D$ for Ag/SnS device. Each data points of SHG and ON/OFF ration represent an individual SnS crystal. **c** Cross-sectional HAADF-STEM image of 16L SnS along the armchair direction. The scale bars represent 1 nm.

**Supplementary Figures**

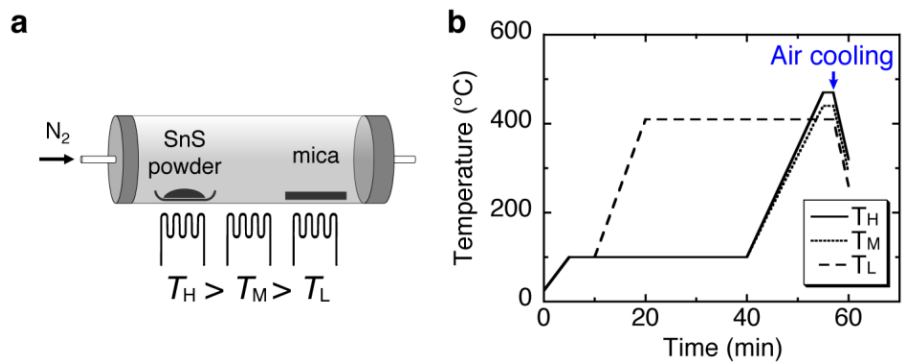

**Supplementary Fig. 1** Growth conditions for monolayer SnS. **a** Schematic diagram of three-zone PVD chamber by separately controlling the heaters at high, middle, and low temperatures ($T_H$, $T_M$, and $T_L$) **b** Typical temperature profiles for monolayer growth. After the growth, the chamber was air-cooled at the rate of ~50 °C/min.

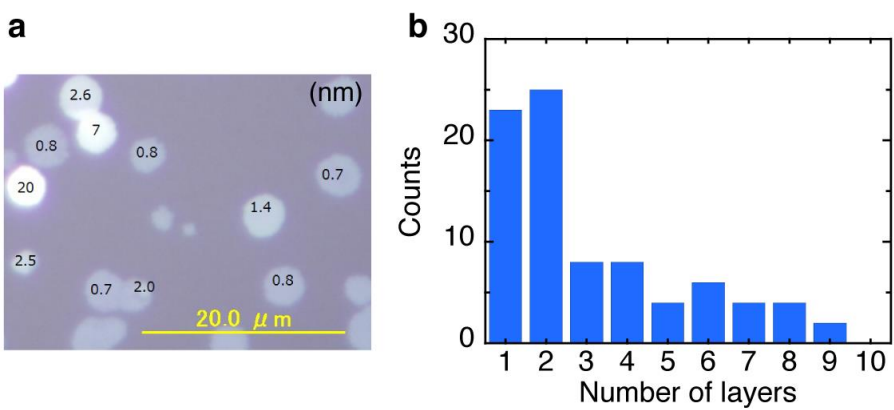

**Supplementary Fig. 2** Thickness distribution of PVD grown SnS. **a** Typical optical image and **b** thickness distribution histogram of SnS on mica with different thickness obtained via the same substrate and growth conditions (Supplementary Fig. 1b).

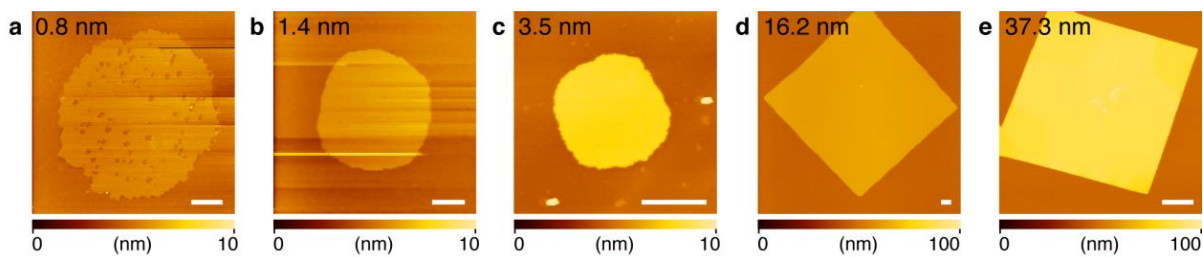

**Supplementary Fig. 3** Evolution of crystal morphology for different SnS thicknesses. **a** 0.8 nm, **b** 1.4 nm, **c** 3.5 nm, **d** 16.2 nm, **e** 37.3 nm. The scale bars represent 1 µm.

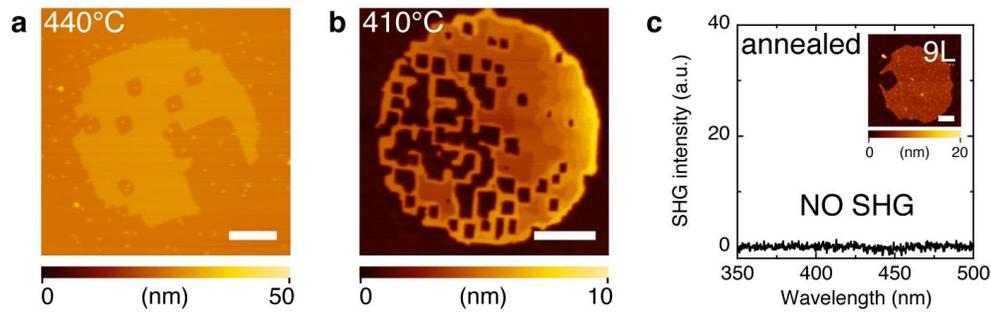

**Supplementary Fig. 4** Effects of post-growth annealing on surface morphology and SHG property of SnS: AFM topographic images of multilayer SnS after post-growth annealing in $N_2$ atmosphere at **a** 440°C and **b** 410°C. Square-shaped etch pits were clearly observed within the SnS crystals, and their lateral size becomes larger with increasing annealing temperature. **c** SHG measurement for 9L SnS after the annealing at 410°C. As shown in the inset AFM image, a square-shaped etch pit were formed near the crystal edge. Even though no other etch pits were found in much of the SnS surface region, SHG signal was undetectable for the annealed sample probably due to the introduction of extrinsic defects. This result confirms that non-centrosymmetry and ferroelectricity are very sensitive to the crystalline quality. The scale bars represent 500 nm.

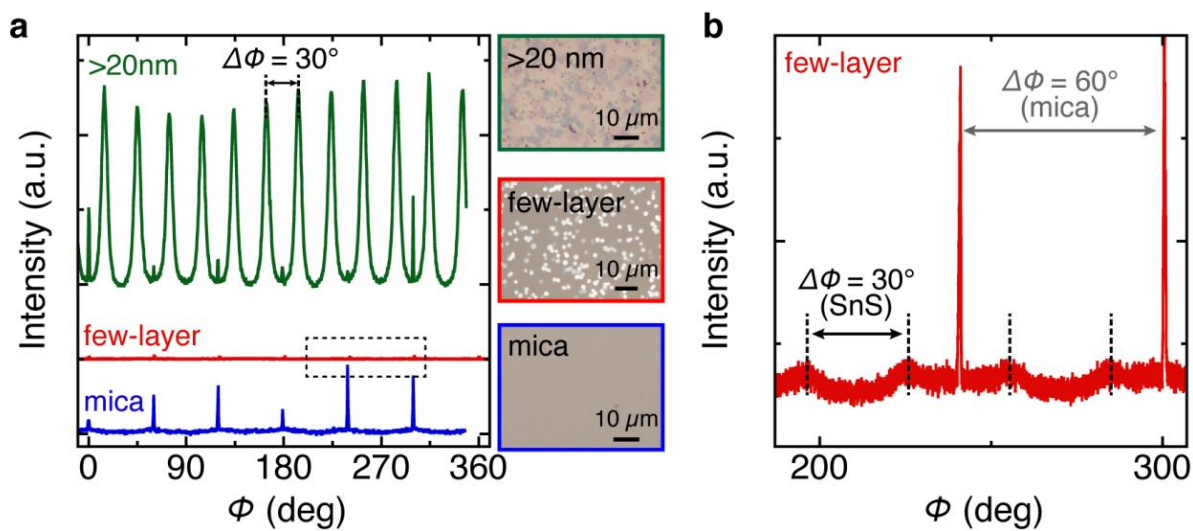

**Supplementary Fig. 5** XRD analysis of SnS grown on mica. **a** In-plane XRD Φ-scan for PVD grown SnS on mica measured with the samples tilted at an angle of $\chi = 23.3°$, which corresponds to the angle between SnS(016) and (001) planes[39]. The growth time was changed from 2 to 15 min. The 15-min grown sample was full covered with >20 nm SnS crystals, whereas the 2-min grown sample has small coverage with isolated multi-layer SnS crystals. As a reference, the Φ-scan of intrinsic mica substrate is shown at the bottom. For mica, six peaks were observed with an interval of 60°. The fluctuation of diffraction intensity is probably owing to an inhomogeneous distribution of SnS crystals on mica substrate. **b** Φ-scan for few-layer SnS with the selected region as shown in **a**. The Φ-scan of SnS(016) plane indicates 12 peaks with an interval of 30° for few-layer and >20 nm SnS, which agrees well with the reported data for SnS on mica[39]. These results indicate a strong interaction between SnS and mica probably due to a lattice matching. Although the crystallographic interaction between SnS/mica may disappear with increasing the SnS thickness, there is still a preferred orientation for SnS growth as long as the thickness is several-tens nanometers.

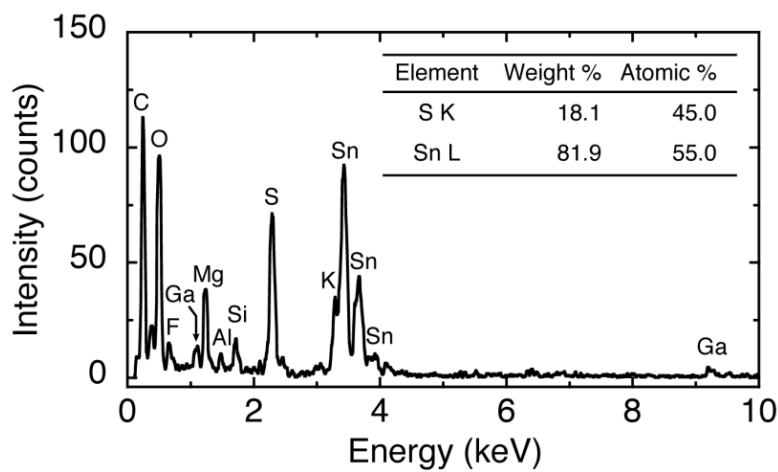

**Supplementary Fig. 6** Typical cross-sectional EDS spectrum of trilayer SnS.

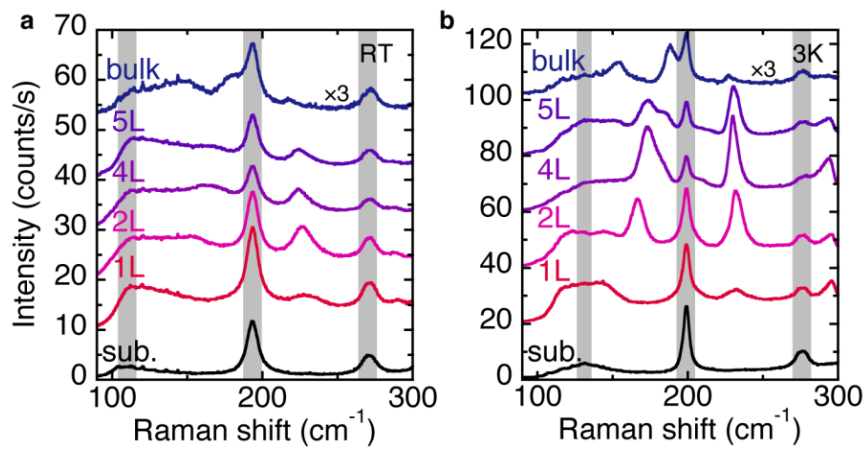

**Supplementary Fig. 7** Typical Raman spectra for SnS with different thicknesses from bulk to monolayer, measured at **a** RT and **b** 3K.

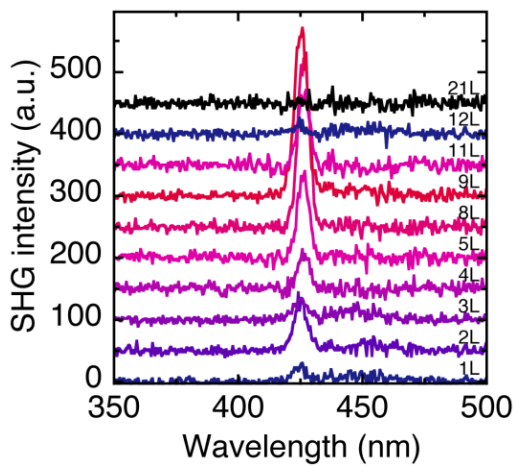

**Supplementary Fig. 8** Typical RT SHG spectra for SnS with different thicknesses.

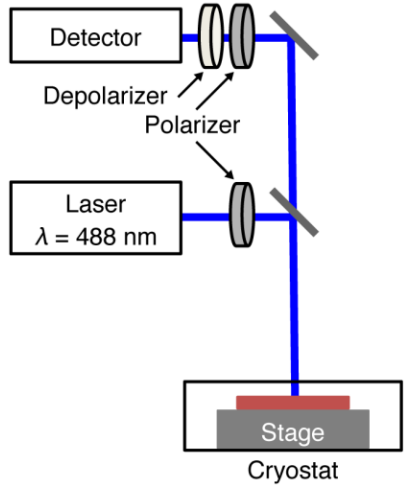

**Supplementary Fig. 9** Schematic diagram of polarized Raman spectroscopy. The polarization angle was changed through rotating the polarizers. The second polarizer was set to be parallel or perpendicular to the polarized incident light for Raman measurement.

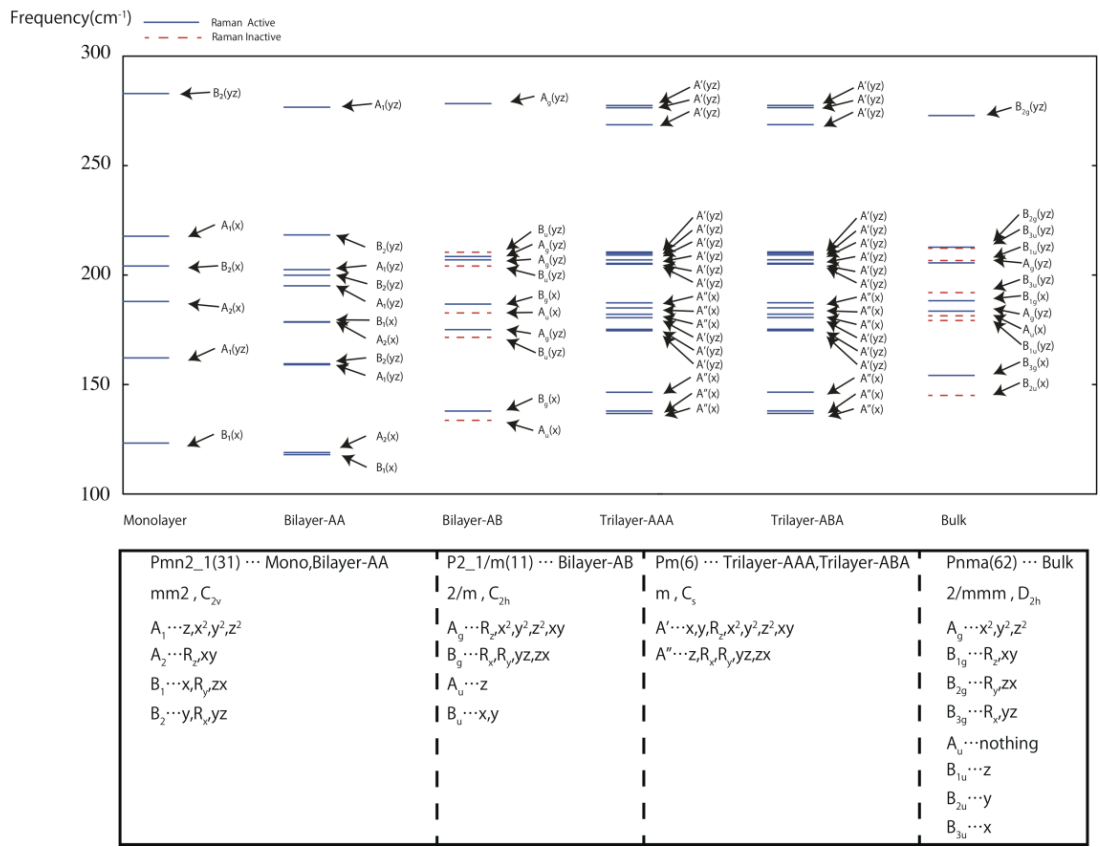

**Supplementary Fig. 10** Calculated Raman active/inactive phonon modes for monolayer, AA/AB-bilayer, AA/AB-trilayer, and AB-bulk SnS. The point group, irreducible representation of phonon modes, and vibration direction are summarized.

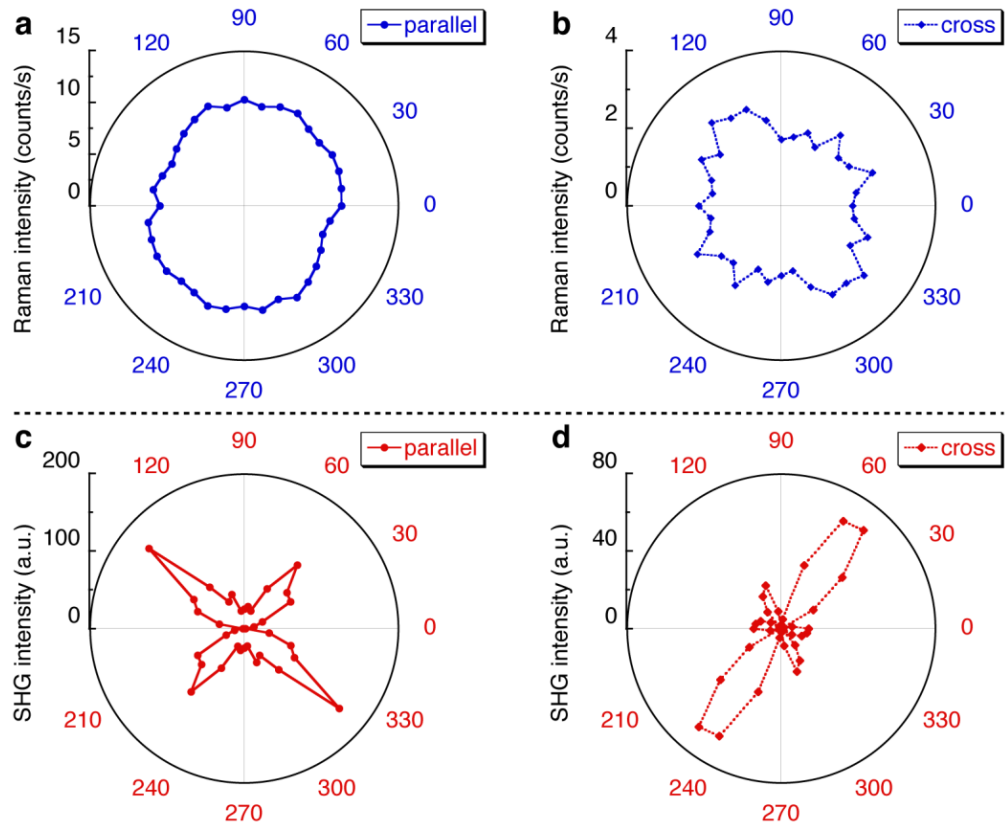

**Supplementary Fig. 11** Angular dependence of optical measurements for 10L SnS: Raman intensity at ~226 cm$^{-1}$ under **a** parallel and **b** perpendicular polarization, and SHG intensity under **c** parallel and **d** perpendicular polarization. Note that different 10L SnS samples were used for each measurement. A strong anisotropy was confirmed for the SHG intensity, though angular dependence were slight for the Raman measurements. This discrepancy is probably due to the difference of crystalline quality, which also resulted in the large distribution of SHG intensity even for the same thickness (**Fig. 5b**).

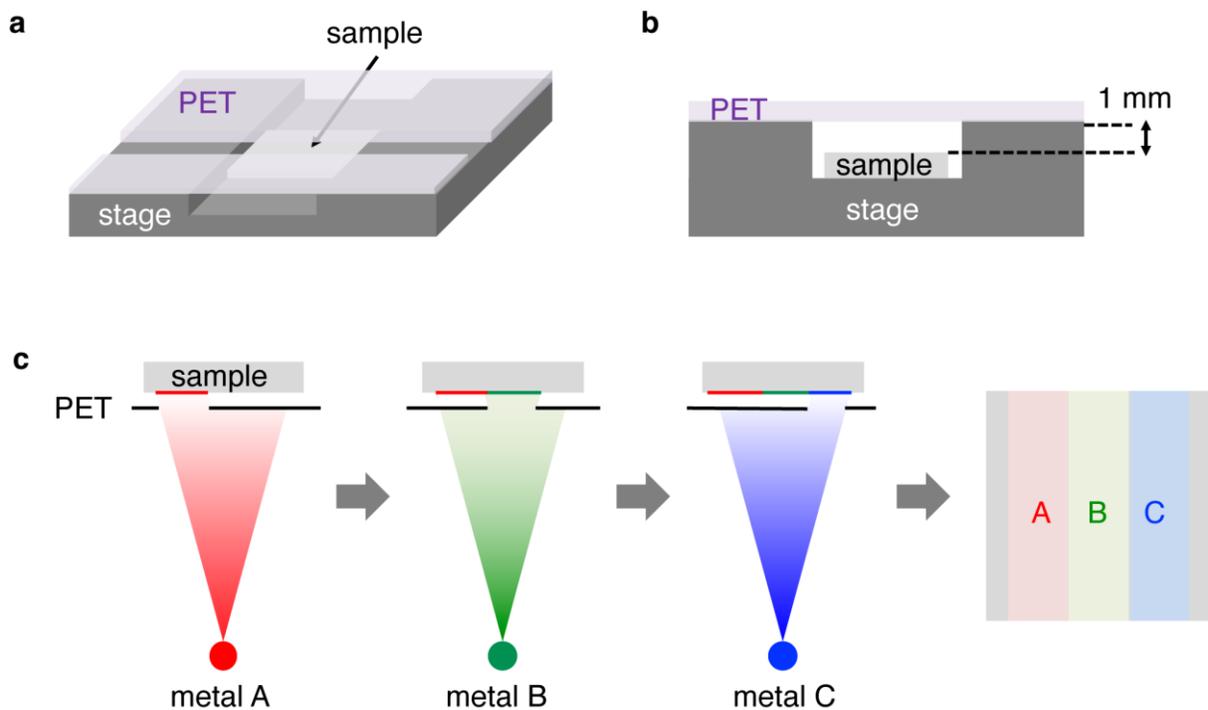

**Supplementary Fig. 12** Fabrication method of multiple metal depositions. **a** Bird's-eye view and **b** cross-sectional view of sample stage with PET shield. **c** Step-by-step deposition of multiple metals.

**Method:** After the EB lithography for the electrodes, the sample was set on a home-made stage with hollow (**a,b**). The sample surface was partially covered with the PET shield, which was placed approximately 1 mm apart from the sample. Each metal was deposited step-by-step with changing the position of the PET shield by using a standard thermal evaporator (**c**). Finally, SnS devices with different electrode metals were obtained on the same wafer, as shown in Fig. 4a.

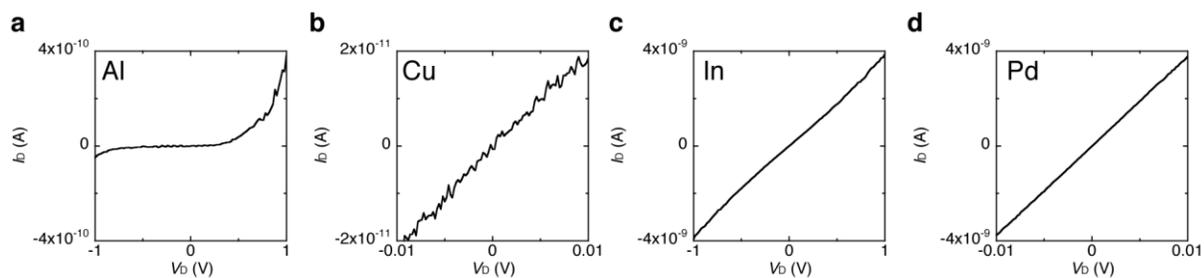

**Supplementary Fig. 13** Typical $I_D$–$V_D$ curves for bulk SnS with different contact metals: **a** Al, **b** Cu, **c** In, and **d** Pd.

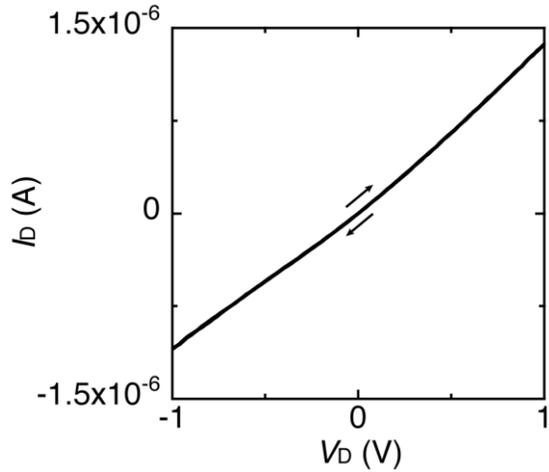

**Supplementary Fig. 14** Typical $I_D$–$V_D$ curve with Ni contact for few-layer SnS below the critical thickness for ferroelectricity.

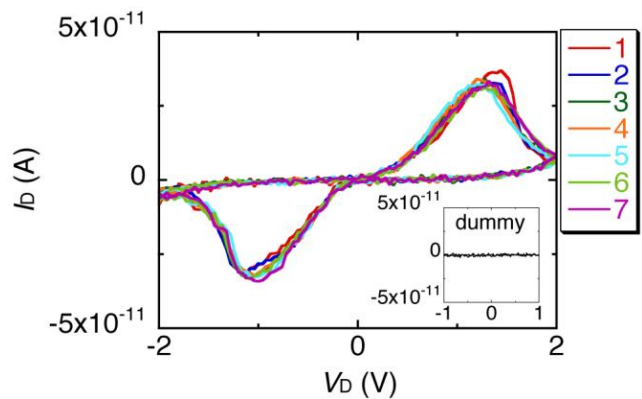

**Supplementary Fig. 15** Ferroelectric switching characteristic of $I_D$–$V_D$ for Ag/9L-SnS cycled 7 times at RT. Inset: $I_D$–$V_D$ for a dummy sample without SnS. Ferroelectric hysteresis was absent for the dummy device without SnS. The reproducible switching characteristic was observed in the multiple measurements, indicating the stability of ferroelectricity.

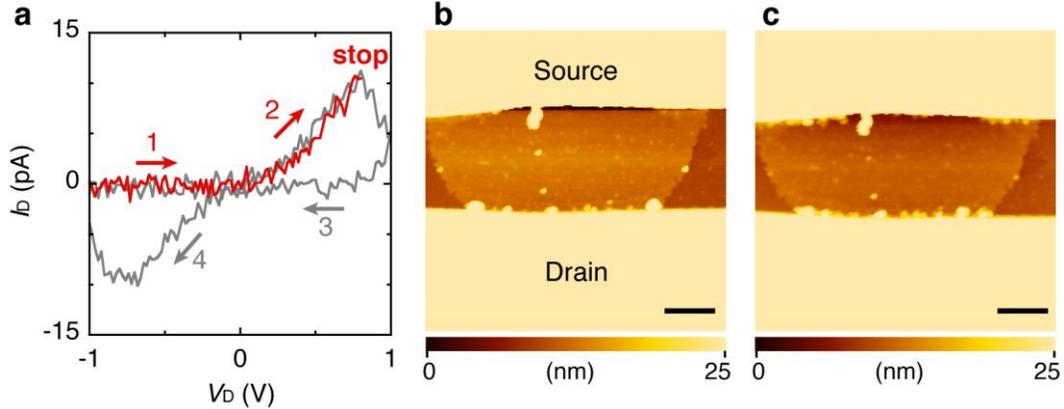

**Supplementary Fig. 16** Exclusion of Ag diffusion into SnS. **a** $I_D$–$V_D$ curves for 2.6 nm (~4L) SnS with Ag electrodes. A single sweep from −1 V to 800 mV stopped at the low resistive state is shown in a red line. For comparison, a double sweep is shown in a gray line. AFM topographic images **a** before and **b** after applying the external electric field. The scale bars represent 500 nm.

AFM topographic image was observed after the $I_D$–$V_D$ measurement. In order to observe the SnS surface at the low resistive state (LRS), the drain voltage was swept from −1 V to 800 mV just before the AFM measurement (a). No significant Ag diffusion was confirmed in the channel region at the LRS compared with the AFM image before the electrical measurement (b,c).

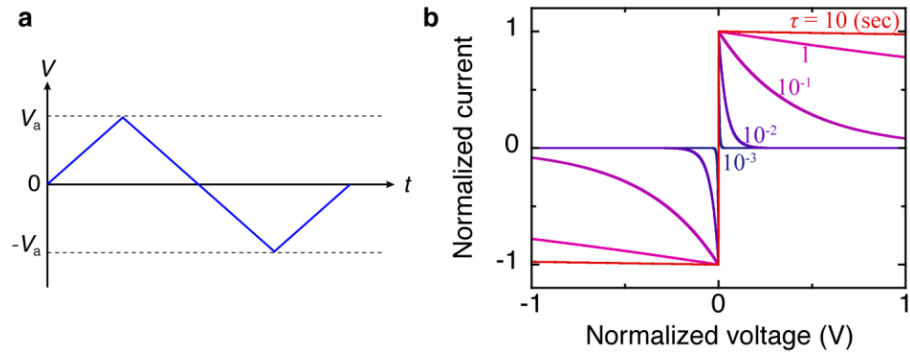

**Supplementary Fig. 17** Theoretical analysis of *I–V* hysteresis loop based on the carrier trapping model. **a** The schematic diagram of applied voltage. **b** Calculated relationship between voltage and current from carrier emission. The emission time constant $\tau$ was changed from 1 ms to 10 s.

**Analysis:** False ferroelectricity in nonferroelectric systems has been theoretically investigated based on the carrier trapping model[55]. In this model, the following properties are assumed; 1) the sample is a semiconductor with Schottky contacts on both the source and drain contacts, 2) a large concentration of trap sites is located over a finite thickness $w_t$ at the semiconductor/metal interfaces, 3) the normal leakage current is neglected and the carrier emission from the traps is dominant, and 4) the displacement current due to the space charge near the contacts produced by the trapping/detrapping is neglected. Based on this model, the carrier emission from the traps $I_{tr}$ can be written as

$$I_{tr} = \frac{qAw_tN}{\tau}\exp\left(-\frac{V}{4\tau V_a f}\right) \qquad (S1)$$

where $A$ is the area of the electrode, $N$ is the initial concentration of occupied traps, $\tau$ is the emission time constant from the traps, $V$ is the applied source–drain voltage, $V_a$ is the amplitude of the applied signal (Supplementary Fig. 13a), and $f$ is its frequency.

To exclude the possibility that the present ferroelectric switching behavior is due to the carrier emission from the traps, we discuss the suitability of carrier trapping models assuming that the present device is not ferroelectric. When the relationship of current and voltage is calculated based on Eq. (S1), the shape of the $I$–$V$ curve strongly depends on $\tau$, as shown in Supplementary Fig. 13b. Comparing the $I$–$V$ shapes of the calculated and experimental results (Figs. 4c,e and 5a), the value of $\tau$ is approximately estimated to be in the order of 100 ms. The maximum current $I_{max}$ can be obtained when the voltage is close to zero. Substituting $V = 0$ V into Eq. (S1), the total number of traps $Aw_tN$ can be expressed as follows,

$$Aw_tN = \frac{\tau}{q}I_{max} \qquad (S2)$$

From the experimental results, $I_{max}$ is ranged up to 150 pA (Fig. 4e). When we substitute $\tau = 100$ ms and $I_{max} = 150$ pA, the total number of traps is determined to be $9 \times 10^7$ states near the Ag/SnS interface in the SnS channel (the channel thickness, length, and width are 5.4 nm, 0.4 µm, and 3.0 µm, respectively). In this case, the number of traps corresponds to one in three of the total number of atoms in the SnS channel, which is too large intuitively. From these results, the effect of carrier emission from the traps can be neglected for the present ferroelectric $I$–$V$ hysteresis loops.

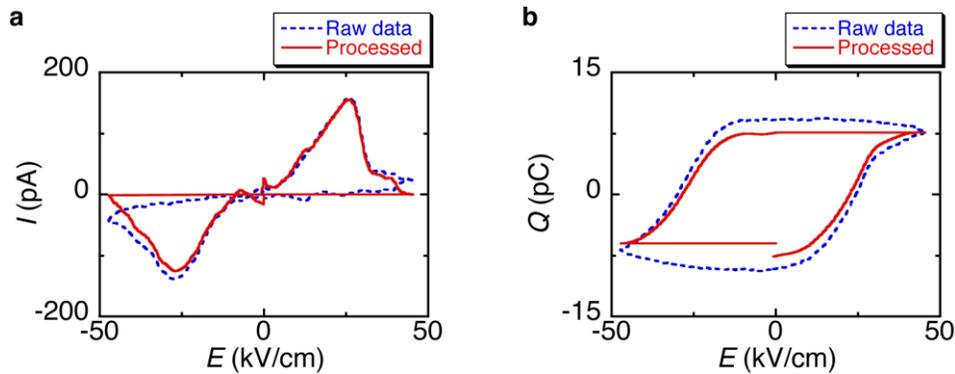

**Supplementary Fig. 18** Background subtraction for $I$–$E$ and $Q$–$E$ hysteresis loop. **a** To eliminate the residual contribution of leakage current in $I$–$E$ hysteresis loop, the current value at the high resistive state was subtracted from that at the low resistive state. **b** $Q$–$E$ hysteresis loop with and without the background subtraction.

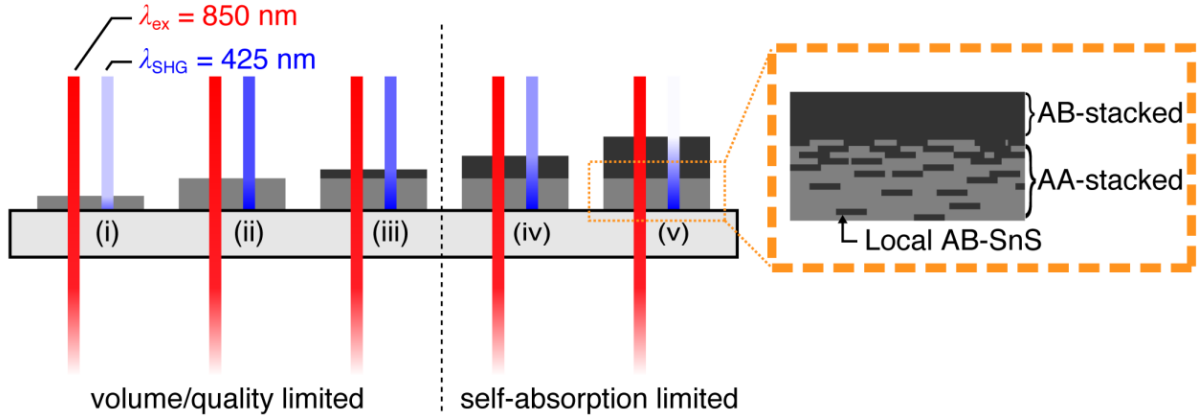

**Supplementary Fig. 19** Schematic illustration of a model for interlayer coupling in multilayer SnS. The red and blue lines indicate the excitation laser ($\lambda$ = 850 nm) and SHG signal ($\lambda$ = 425 nm), respectively.

In this model, for the thin SnS crystals, the stacking sequence is dominated by AA stacking probably due to the effect of strain introduced through the interaction with the mica substrate. In contrast, above the critical thickness of ~15L, the stacking sequence is gradually changed to AB stacking, which is a thermodynamically stable state. In the SHG measurements, an 850-nm excitation laser, whose penetration depth in SnS is at least of several-hundred-nm[60,61], was used to generate the SH signals. Thus, in the case of SnS thinner than ~20L as discussed in this work, the whole region is excited. When the SnS thickness reaches the critical thickness at (iii), the SHG signal is maximized. Above the critical thickness, although the volume of non-centrosymmetric region does not change, the self-absorption of the SHG signal at 425 nm becomes effective because its penetration depth is much shorter than 850 nm[60,61]. As a result, the SHG intensity decreases with increasing SnS thickness and is finally annihilated.

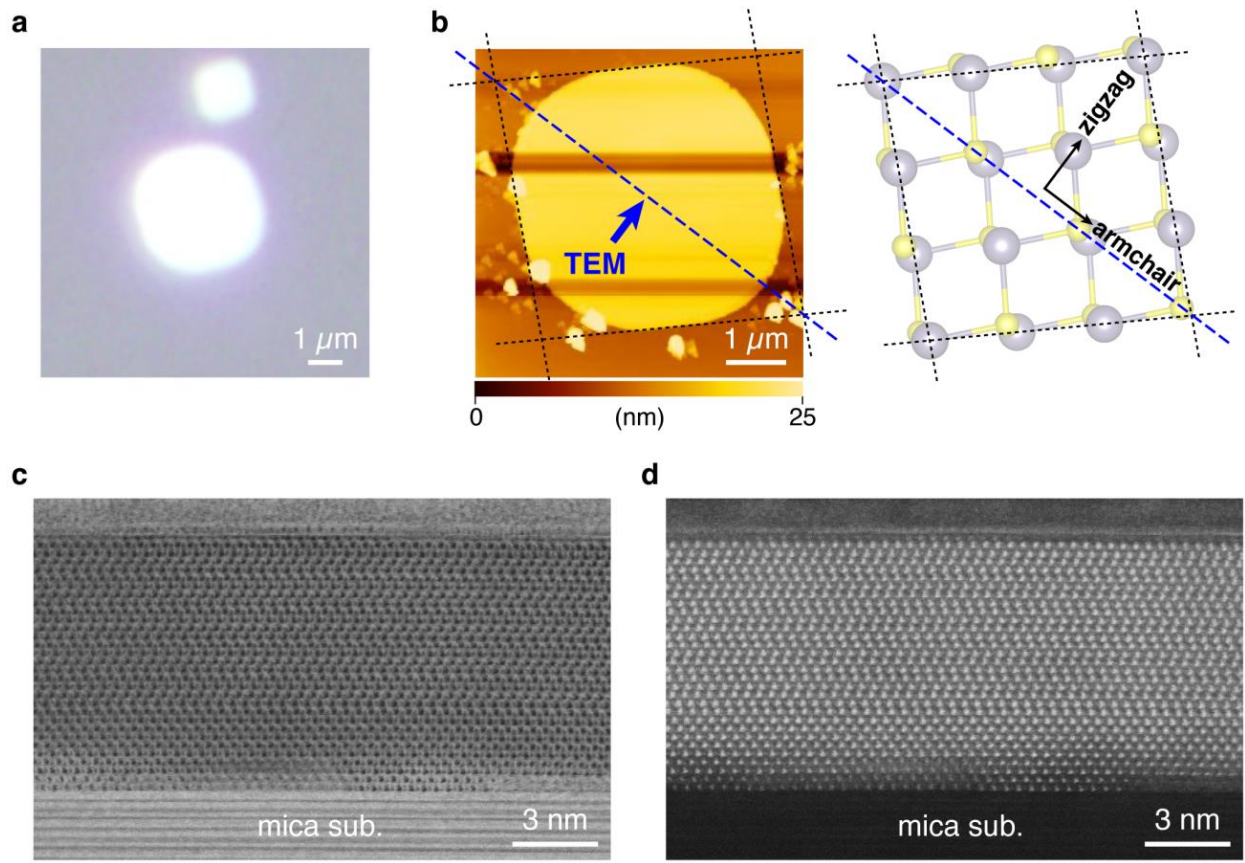

**Supplementary Fig. 20** Cross-sectional TEM observation of PVD grown SnS on mica. **a** Optical and **b** AFM topographic images of 16L SnS. The sample was cut along the armchair direction, as shown in a dashed blue line. **c** Bright-field STEM and **d** HAADF-STEM images of 16L SnS.

**Supplementary References**